\documentclass[12pt]{article}
\usepackage{graphicx,color}
\usepackage{booktabs,multirow}
\usepackage{amsmath,XoohmE}

\def\df{\dot\phi}
\def\dj{\dot\psi}

\definecolor{Hey}{rgb}{.9,.05,.4}
\definecolor{plum}{rgb}{.4,0,.6}
\definecolor{Tan}{rgb}{.87,.62,0.42}
\definecolor{Green}{rgb}{0.25,.65,.25}
\definecolor{Plum}{rgb}{.47,0.15,.7}
\definecolor{Red}{rgb}{.75,0.15,0.25}
\definecolor{Cyan}{rgb}{0.4,0.6,.95}
\def\TE{\hbox{\color{Tan}\boldmath$\e$}^1}
\def\GE{\hbox{\color{Green}\boldmath$\e$}^2}
\def\PE{\hbox{\color{Plum}\boldmath$\e$}^3}
\def\RE{\hbox{\color{Red}\boldmath$\e$}^4}
\def\CE{\hbox{\color{Cyan}\boldmath$\e$}^5}

\def\vC#1{\vcenter{\hbox{\hss#1\hss}}}
\newtheorem{corr}{Correspondence}

\def\Ft#1{\,\footnote{#1}}
\newdimen\parshift\parshift=\parindent
\catcode`@=11
 \long\def\@footnotetext#1{\insert\footins{\reset@font\footnotesize
           \interlinepenalty\interfootnotelinepenalty\splittopskip%
            \footnotesep\splitmaxdepth\dp\strutbox\floatingpenalty\@MM%
             \hsize\columnwidth\addtolength{\hsize}{-2\parindent}
              \@parboxrestore\protected@edef\@currentlabel%
              {\csname p@footnote\endcsname\@thefnmark}%
                \color@begingroup\baselineskip=10pt%
                 \@makefntext{\rule\z@\footnotesep\ignorespaces#1%
                  \@finalstrut\strutbox}%
                \color@endgroup}}
 \long\def\@makefntext#1{\hglue\parshift%
           \vbox{\noindent\hb@xt@0em{\hss\@makefnmark\kern1pt}#1}}
\catcode`@=12
 \def\ba{\left(\begin{array}}
 \def\ea{\end{array}\right)}

 \def\brr{\begin{eqnarray}}
 \def\err{\end{eqnarray}}

 \font\rOpe=cmsy10                        
 \def\ktl{{\hbox{\rOpe\char'170}}}        
 \def\kbl{{\hbox{\rOpe\char'170}}}        
 \def\kcr{{\reflectbox{\rOpe\char'170}}}  
 \def\ktr{{\reflectbox{\rOpe\char'170}}}  
 \def\kbr{{\reflectbox{\rOpe\char'170}}}  
 \def\Border{\vbox{\hsize0pt
        \setlength{\unitlength}{1mm}
        \newcount\xco
        \newcount\yco
        \xco=-21
        \yco=12
        \begin{picture}(0,0)(-7.5,0)
        \put(\xco,\yco){$\ktl$}
        \advance\yco by-1
        {\loop
        \put(\xco,\yco){$\kcr$}
        \advance\yco by-2
        \ifnum\yco>-240
        \repeat
        \put(\xco,\yco){$\kbl$}}
        \xco=170
        \yco=12
        \put(\xco,\yco){$\ktr$}
        \advance\yco by-1
        {\loop
        \put(\xco,\yco){$\kcr$}
        \advance\yco by-2
        \ifnum\yco>-240
        \repeat
        \put(\xco,\yco){$\kbr$}}
        \put(-19.5,13){\scalebox{.54}{State University of New York
            Physics Department|University of Maryland Center for
            String and Particle  Theory \&\ Physics Department|%
            Howard University Physics \&\ Astronomy Department}}
        \put(-19.5,-241.5){\scalebox{.69}{University of Washington
            Mathematics Department|Pepperdine University Natural
            Sciences Division|University of Oregon Mathematics
            Department}}
        \end{picture}
        \par\vskip-8mm}}
\definecolor{UMred}{rgb}{.9,.05,.2}
 \def\UMbanner{\vbox{\hsize0pt
        \setlength{\unitlength}{.4mm}
        \thicklines
        \begin{picture}(0,0)(-30,-10)
        \put(165,16){\line(1,0){4}}
        \put(170,16){\line(1,0){4}}
        \put(180,16){\line(1,0){4}}
        \put(175,0){\line(1,0){4}}
        \put(180,0){\line(1,0){4}}
        \put(185,0){\line(1,0){4}}
        \put(169,0){\line(0,1){16}}
        \put(170,0){\line(0,1){16}}
        \put(179,0){\line(0,1){16}}
        \put(180,0){\line(0,1){16}}
        \put(184,0){\line(0,1){16}}
        \put(185,0){\line(0,1){16}}
        \put(169,16){\oval(8,32)[bl]}
        \put(170,16){\oval(8,32)[br]}
        \put(179,0){\oval(8,32)[tl]}
        \put(185,0){\oval(8,32)[tr]}
        \end{picture}
        \par\vskip-6.5mm
        \thicklines}}

 \SfTitles 
 \allowdisplaybreaks

\begin{document}
\thispagestyle{empty}
\vbox{\Border\UMbanner}
 \noindent
 \today\hfill{
              UMDEPP 06-056 
 }
 \begin{center}
{\LARGE\sf\bfseries\boldmath
  A Counter-Example to a Putative Classification\\[2mm]
  of 1-Dimensional, $N$-extended Supermultiplets}\\[5mm]
{\sf\bfseries C.F.\,Doran$^a$, M.G.\,Faux$^b$, S.J.\,Gates, Jr.$^c$,
     T.\,H\"{u}bsch$^d$, K.M.\,Iga$^e$ and G.D.\,Landweber$^f$}\\[2mm]
{\small\it
  \leavevmode\hbox to\hsize{\hss
  $^a$Department of Mathematics,
      University of Washington, Seattle, WA 98105%
  , {\tt  doran@math.washington.edu}\hss}
  \\
  $^b$Department of Physics,
      State University of New York, Oneonta, NY 13825%
  , {\tt  fauxmg@oneonta.edu}
  \\
  $^c$Center for String and Particle Theory,\\[-1mm]
      Department of Physics, University of Maryland, College Park, MD 20472%
  , {\tt  gatess@wam.umd.edu}
  \\
  $^d$Department of Physics \&\ Astronomy,
      Howard University, Washington, DC 20059%
  , {\tt  thubsch@howard.edu}\\[-1mm]
  \leavevmode\hbox to\hsize{\hss
  Department of Applied Mathematics and Theoretical Physics,
      Delaware State University, Dover, DE 19901\hss}
  \\
  $^e$Natural Science Division,
      Pepperdine University, Malibu, CA 90263%
  , {\tt  Kevin.Iga@pepperdine.edu}
  \\
 $^f$Mathematics Department,
     University of Oregon, Eugene, OR 97403-1222%
  , {\tt  greg@math.uoregon.edu}
 }\\[3mm]
{\sf\bfseries ABSTRACT}\\[3mm]
\parbox{5.3in}{
We present a counter-example to a recent claim that supermultiplets of $N$-extended supersymmetry with no central charge and in 1-dimension are specified unambiguously by providing the numbers of component fields in all available engineering dimensions within the supermultiplet.}

\end{center}
\begin{flushright}\sl
 Thought is impossible without an image.\\*[-1mm]
 |~Aristotle
\end{flushright}

\section{Introduction, Results and Summary}
 \label{IRS}
 Supersymmetric systems have been studied over more than three decades and find many applications, although experimental evidence that Nature also employs supersymmetry is ironically lacking within high-energy particle physics wherein it was originally invented. Nevertheless, supersymmetry is also a keystone in most contemporary attempts at unifying all fundamental physics, such as string theory and its $M$- and $F$-theory extensions. There, one typically needs a large number, $N\leq32$, of supersymmetry generators, in which case {\em off-shell\/} descriptions, indispensable for a full understanding of the quantum theory, are sorely absent.

 This has motivated some of the recent interest%
\cite{rGR0,rGR1,rGR2,rKRT,rT06,rGLPR,rGLP} in the 1-dimensional dimensional reduction of supersymmetric field theories\Ft{We refer to $N$-extended supersymmetry in 1-dimensional time as ``$(1|N)$-supersymmetry''. Besides dimensional reduction of field theories in higher-dimensional spacetime to their 1-dimensional shadows, $(1|N)$-supersymmetry is also present in the study of supersymmetric wave functionals in any supersymmetric quantum field theory and so applies to all of them also in this other, more fundamental way.}, where subtleties stemming from the Lorenz groups in spacetimes of various dimensions can be deferred, to be incorporated subsequently, while reconstructing the original higher-dimensional theory; cf.\ the {\sc radio} method of Ref.\cite{rGR0}. 
 In particular, Refs.\cite{rA,r6-1,r6-2} introduce, hone and apply a graphical device, akin to wiring schematics, which can fully encode all requisite details about $(1|N)$-supersymmetry, its action within off-shell supermultiplets and the possible couplings of all off-shell supermultiplets---for all $N$. These graphs, called {\em Adinkras\/}, are closely related to the rigorous underpinning of off-shell representation theory in supersymmetry\cite{r6--1}, but are also intuitively easy to understand and manipulate. Finally, where a superfield realization of the supermultiplets is known, translation between these and Adinkras is straightforward\cite{r6-1}.
 
 This combination of precision, exactness and intuitive ease makes Adinkras a natural tool for exploring the sometimes unexpected intricacies in the supersymmetric zoo, much as Feynman diagrams facilitate many-body and quantum field theory computations.
 \ping
 
 Adinkras represent component bosons in a supermultiplet as white nodes, and fermions as black nodes. A white and a black node are connected by an edge, drawn in the $I^{\rm th}$ color if the $I^{\rm th}$ supersymmetry transforms the corresponding component fields one into another. A sign/parity degree of freedom\cite{rA} in the supersymmetry transformation of a component field into another is represented by solid {\em vs\/}.\ dashed edges.
 In the natural units ($\hbar=1=c$), all physical fields have a definite engineering dimension, defined up to an overall additive constant, and we accordingly stack the nodes at heights that reflect the engineering dimensions of the corresponding component fields.
 With this implicit upward orientation of the edges, in the direction of increasing engineering dimension, a node is a {\em source\/} if no lower node connects to it, and a {\em target\/} if no higher node connects to it. We then have:
 \begin{corr}\label{c:A2F}
 A target in an Adinkra corresponds to a component field the supersymmetry transform of which contains only time-derivatives of other component fields.
 A source in an Adinkra corresponds to a component field the supersymmetry transform of which contains no time-derivatives of other component fields.
\end{corr}
Finally, there exists a precise 1--1 dictionary between the various graphical characteristics of the elements in Adinkras and relations between them to the concrete properties of the corresponding component fields and relations between them in the corresponding supermultiplet\cite{rA,r6-1}. As a matter of principle and to showcase the inherent power of this graphical tool, we defer the familiar notation of supersymmetry transformations to the appendix: `an Adinkra is worth ten thousand equations.'
\ping

 The purpose of this short note is to demonstrate the ease of use of Adinkras by presenting a counter-example to the claim\cite{rKRT,rT06} that $(1|N)$-supermultiplets are specified unambiguously by providing the numbers of component fields in all available engineering dimensions within the supermultiplet.

The simplest way to understand the failing of the putative classification scheme in\cite{rKRT} is to note an analogy. In Lie algebra theory, it is well known that the classification of the dimensionality of irreps is {\it {not}} equivalent to a complete classification of irreps. So for example, within $\mathfrak{su}(3)$, there are {\it two} inequivalent 20-dimensional irreps. While the work in\cite{rKRT} claims to classify supersymmetry irreps, in fact at most it only classifies the analog of the dimensionality of the supersymmetry irreps.

\section{The Counter-Example}
 \label{s:CE}
Refs.\cite{rKRT,rT06} imply\Ft{\label{f:Exp}We find it hard to pinpoint {\em what\/} Ref.\cite{rKRT} in fact does claim: In appendix~A, case~{\it vi}), they write: ``The length 2 and 3 irreps are obtained from the $N=8$ irreps by restricting the supersymmetry transformations to be given by $Q_i$, for $i=1,\cdots,5$.'' This is trivially true of all $N<5$ supermultiplets. However, we find that depending on the number of distinct engineering dimensions available in a supermultiplet, the result may or may not be unique, depending on which $Q_i$'s were omitted. Without this additional information, the statement seems rather vacuous. In turn, case {\it vii}) of the same appendix presents a list of ``quaternionic $N=5$ irreps'', which has single $(6,8,2)$- and $(7,8,1)$-entries, whereas\eq{e682w2} and\eq{e682w4} are clearly inequivalent, while the $(7,8,1)$-dimensional one in Figure~\ref{f:781-682} is unique. The Authors of Ref.\cite{rKRT} disagree that this implies either an ambiguity or an incompleteness in their classification; may the Reader decide.} that $(1|N)$-supermultiplets without central charge are classified by sequences of integers such as $(n_0,n_1,\cdots)$, where $n_k$ denotes the number of component fields of engineering dimension $d_0+k/2$; $d_0$ is constant for each $(1|N)$-supermultiplet and irrelevant for our purposes.

In particular, the results of Refs.\cite{rKRT,rT06} thus imply that there is, \eg, a single $(6,8,2)$-dimensional $(1|5)$-supermultiplet\Ft{Ref.\cite{rKRT} in fact lists {\em two\/} $N=5$ irreps of every graded dimension, but says nothing about (in)equivalence by field redefinitions between matching pairs. Were they intended as {\em inequivalent\/}, it is the unique $(7,8,1)$-dimensional supermultiplet given in Fig.~1, below, that is a counter-example. In any case, however, the existence of {\em two\/} $N=6$ supermultiplets\eqs{e682x4}{e682x6} contradicts Ref.\cite{rKRT}.}. Not so: both
\begin{equation}
 \vC{ \begin{picture}(135,50)(0,0)
  \put(-10,2){\includegraphics[height=50mm]{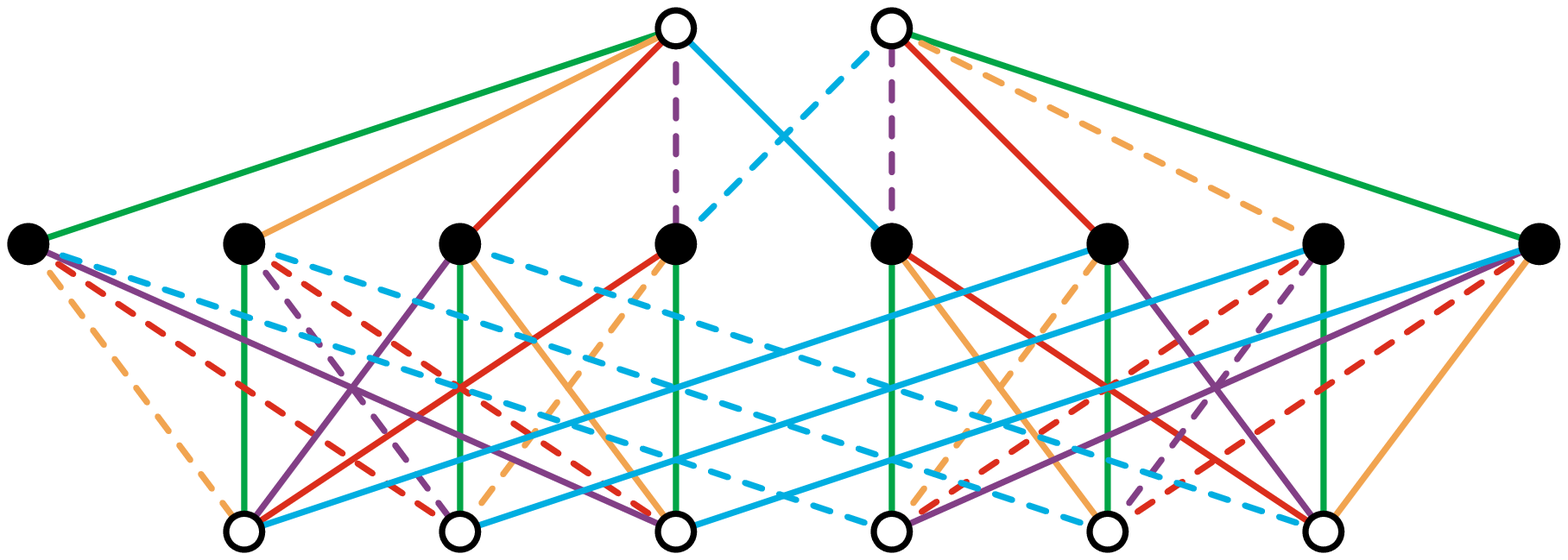}}
  \put(7,2){1}
  \put(26.75,2){2}
  \put(46.5,2){3}
  \put(74,2){4}
  \put(93.5,2){5}
  \put(113.5,2){6}
  \put(46.5,50){7}
  \put(74,50){8}
  \put(-13,30){1}
  \put(7,30){2}
  \put(27,30){3}
  \put(47,30){4}
  \put(73.5,30){5}
  \put(93.5,30){6}
  \put(113.5,30){7}
  \put(134,30){8}
 \end{picture}}
 \label{e682w2}
\end{equation}
and
\begin{equation}
 \vC{ \begin{picture}(135,50)(0,0)
  \put(-10,2){\includegraphics[height=50mm]{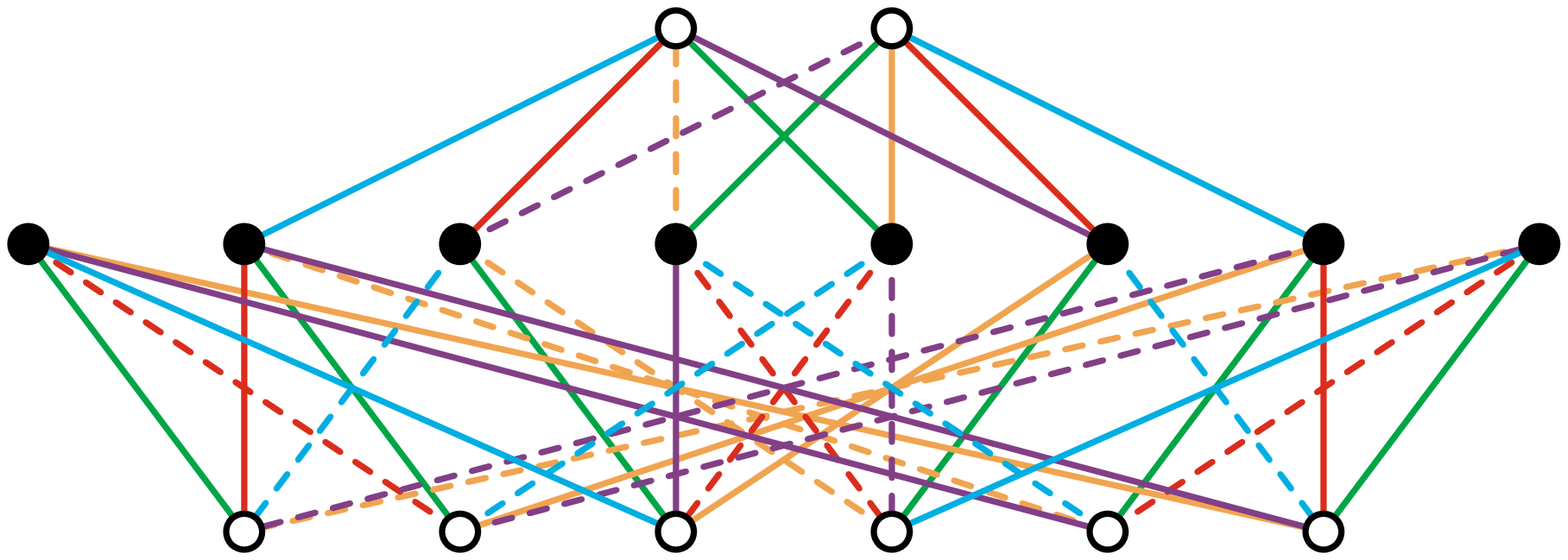}}
  \put(7,2){1}
  \put(26.75,2){2}
  \put(46.5,2){3}
  \put(74,2){4}
  \put(93.5,2){5}
  \put(113.5,2){6}
  \put(46.5,50){7}
  \put(74,50){8}
  \put(-13,30){1}
  \put(7,30){2}
  \put(27,30){3}
  \put(47,30){4}
  \put(73.5,30){5}
  \put(93.5,30){6}
  \put(113.5,30){7}
  \put(134,30){8}
 \end{picture}}
 \label{e682w4}
\end{equation}
represent {\em two distinct\/} $(6,8,2)$-dimensional supermultiplets; see \Eqs{f1}{f8} and\eqs{F1}{F8} in the appendix for the explicit supersymmetry transformation rules. In fact, they are manifestly distinct even at first glance: no amount of horizontal shuffling of the nodes (\ie, basis change of component field of the same engineering dimension) and re-coloring the edges\Ft{We increased the height between the bottom and the middle level a little, to better show the connections.} (\ie, basis change of supersymmetry generators) can turn\eq{e682w2} into\eq{e682w4}. Let $\f_1,\cdots,\f_6,F_7,F_8$ denote the eight component bosons corresponding to the eight accordingly numbered white nodes, and $\j_1,\cdots,\j_8$ the eight component fermions corresponding to the eight accordingly numbered black nodes.
\ping

The distinction between\eq{e682w2} and\eq{e682w4} may be specified in ways that are independent of basis changes of the component fields and/or of the supersymmetry generators: We note that in\eq{e682w2}, only two black nodes (4 and 5) connect, each, to both of the top white nodes (7 and 8).
 By contrast, in\eq{e682w4}, four black nodes (3--6) connect, each, to the top two white nodes (7 and 8).
 So, in the supermultiplet depicted in\eq{e682w2}, the supersymmetry transformation of only two fermions, $\j_4$ and $\j_5$, contain the two top bosons, $F_7$ and $F_8$, which one expects to be auxiliary if $\f_1,\cdots,\f_6$ are dynamical.
  In the supermultiplet depicted in\eq{e682w4}, the supersymmetry transform of each of the four fermions, $\j_3,\cdots,\j_6$, contains the top two bosons, $F_7$ and $F_8$.
 
 While the Adinkras\eqs{e682w2}{e682w4} make this kind of distinction easy to spot by the human eye, there are also other, more easily quantifiable ways to specify the distinction.

Theorem~4.1 and Corollary~4.2 from Ref.\cite{r6-1} ensure that every Adinkra is uniquely specified, respectively, either by its {\em set of targets\/} or by its {\em set of sources\/} and the height assignment of these.
 The Adinkra\eq{e682w2} has two targets ($F_7$ and $F_8$), and six sources ($\f_1,\cdots,\f_6$). The number of sources and targets then are $(6,0,0)$ and $(0,0,2)$, respectively.
 By contrast, in\eq{e682w4} the numbers of sources ($\f_1,\cdots,\f_6$) and targets ($F_7$ and $F_8$, {\em but now also\/} $\j_1$ and $\j_8$) are $(6,0,0)$ and $(0,2,2)$, respectively. As per Theorem~4.1 and Corollary~4.2 in Ref.\cite{r6-1}, either the {\em set\/} of sources and their heights, or the {\em set\/} of targets and their heights uniquely specifies the Adinkra of any given topology (see below). That is, the two Adinkras\eqs{e682w2}{e682w4} both have the same {\em numbers\/}, $(6,0,0)$, of sources at the various heights, \ie, engineering dimensions. However, these {\em sets\/} of nodes are distinguished by how they connect to the rest of the Adinkra: In\eq{e682w2}, none among the source fields, $\f_1,\cdots,\f_6$, connects to a target, whereas the sources in\eq{e682w4}, $\f_1,\cdots,\f_6$, connect to two targets: $\j_1$ and $\j_8$. The formal Correspondence~\ref{c:A2F} translates this into an invariant distinction in the way supersymmetry acts within each of the two supermultiplets corresponding to the Adinkras\eq{e682w2} and\eq{e682w4}.

Finally, already a simple tally of targets in various engineering dimensions, $(0,0,2)$ {\em vs\/}.\ $(0,2,2)$, distinguishes the two Adinkras, \eq{e682w2} and\eq{e682w4}, respectively. This provides an easy numerical quantifier which tells the two supermultiplets\eqs{e682w2}{e682w4} apart; the Reader unconvinced by these depictions of supersymmetry transformations should consult \Eqs{f1}{F8} in the appendix. The formal Correspondence~\ref{c:A2F} easily translates the meaning of this numerical quantifier into invariant (but longer and notationally rather more complex) statements about how the supersymmetry acts within the respective supermultiplets.

\section{The Anatomy of Differences}
 \label{s:ET}
This power of the graphical representation of Adinkras stems from the facts that: (1)~Adinkras faithfully represent all aspects of supersymmetry transformations within any supermultiplet\cite{rA} as is readily seen by comparing\eqs{e682w2}{e682w4} with \Eqs{f1}{F8}, and (2)~they faithfully reflect the structure of the corresponding filtered Clifford supermodules\cite{r6--1}. Deferring to Refs.\cite{rA,r6--1} and forthcoming, more detailed studies\cite{r6-3,r6-5} for the relevant technical proofs, we trust the Reader will be satisfied here with the following, somewhat informal observations:

\paragraph{1}
The {\em topology\/} of an Adinkra is defined by the connectivity of its nodes by its edges, ignoring the height assignments of the nodes and the dashedness of the edges. For example, the two Adinkras\eqs{e682w2}{e682w4} have the same topology. On the other hand, the Adinkras\eq{e682w2} and\eq{e682w4} differ in which two of the eight white nodes are at the top level and which six are at the bottom level. To see this, consider
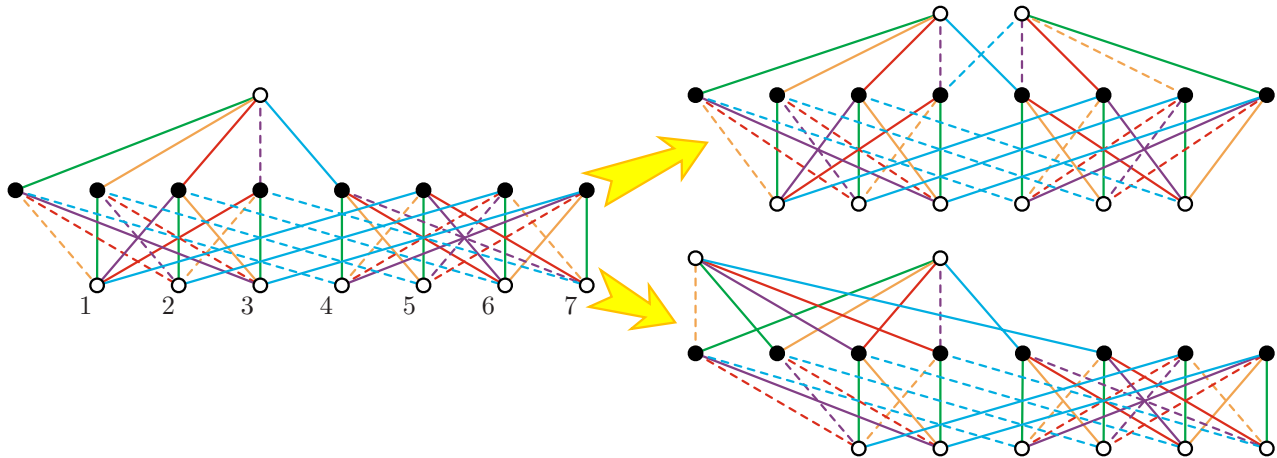
\begin{figure}[htb]
\begin{center}
\begin{center}
 \begin{picture}(140,60)(5,0)
  \put(-8,-2){\includegraphics[height=60mm]{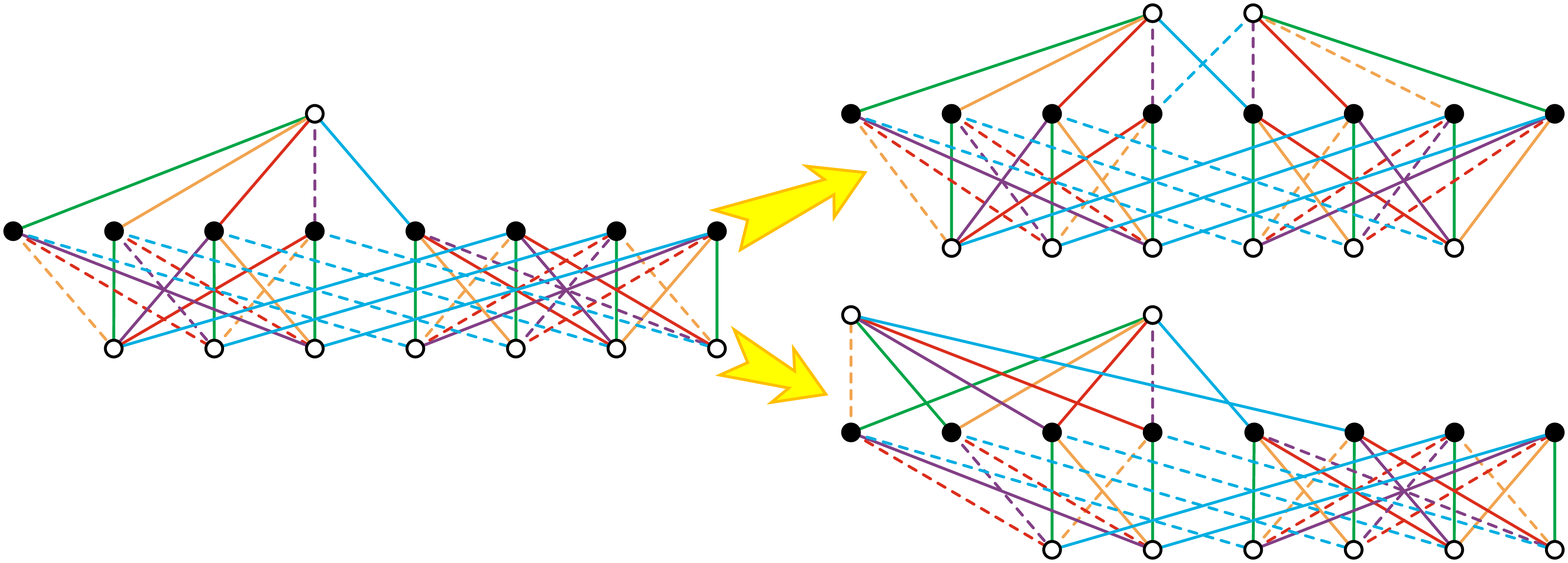}}
   \put(1.5,17){\footnotesize1}
   \put(12.5,17){\footnotesize2}
   \put(23,17){\footnotesize3}
   \put(33.5,17){\footnotesize4}
   \put(44.5,17){\footnotesize5}
   \put(55,17){\footnotesize6}
   \put(66,17){\footnotesize7}
 \end{picture}
\end{center}
\caption{The $(7,8,1)$-dimensional representation of $(1|5)$-supersymmetry with no central charges turns into the two distinct $(6,8,2)$-dimensional representations through vertex-raising\protect\cite{r6-1}---which turns out to equal the application of ``dressing matrices'' in Refs.\protect\cite{rKRT}.}
\label{f:781-682}
\end{center}
\end{figure}
the left-most Adinkra in Fig.~\ref{f:781-682}, which represents a $(7,8,1)$-dimensional $(1|5)$-supermultiplet and which is unique up to a basis change in component fields and/or in supersymmetry generators.
 By raising the white node 7 of the left-most Adinkra, we obtain the top-right one, which equals\eq{e682w2}. Raising instead the nodes 4, 5 or 6 in the left-most Adinkra produces another one that differs from\eq{e682w2} only by horizontally repositioning the nodes and/or re-coloring the edges, \ie, by a basis change of the component fields and/or of the supersymmetry generators.
 On the other hand, by raising node 1 of this left-most Adinkra, we obtain the bottom-right one, which differs from\eq{e682w4} only by some horizontal repositioning of the nodes. Raising instead the nodes 2 or 3 in the left-most Adinkra produces a result that differs from\eq{e682w4} only by a basis change of the component fields and/or of the supersymmetry generators.

\paragraph{2}
 From the appearance and grouping of the nodes in the left-most Adinkra in Fig.~\ref{f:781-682}, it is clear that the two rearrangements to the right are the only two essentially distinct possibilities: the second raised node comes from either the left or the right half of the left-most Adinkra in Fig.~\ref{f:781-682}. The nodes 1, 2 and 3 are each connected to four of the black nodes to which the already raised white node connects; by contrast, the nodes 4--7 are connected to only two of the black nodes to which the already raised white node is connected.
 We reiterate that every statement regarding the connection (or lack thereof) between any two nodes in an Adinkra has an immediate translation into a statement about supersymmetry transforming (or not) a component field into another. The precisely corresponding component field equations are straightforward to write down\cite{rA,r6-1}, but are clearly longer, and notationally rather more complex; see the appendix.

\paragraph{3}
 The `hanging gardens theorems'~5.1--5.4 of Ref.\cite{r6-1} ensure that all Adinkras with the same topology can be obtained from any one of them, by iteratively raising and/or lowering various nodes, which relates to the `automorphic duality' of Refs.\cite{rGR0,rGR1,rGR2,rGLPR}. In forthcoming Ref.\cite{r6-3}, we classify all topologies available to Adinkras corresponding to supermultiplets with no central charge. Jointly, these results classify all $(1|N)$-supermultiplets with no central charge, that are representable by Adinkras. It is fascinating to note that such representations encompass most (if not all) supermultiplets without central charge that have ever occurred in the physics literature, but they by far do not provide for all the representations of such supersymmetry! The task of spelling out the precise circumstances under which a $(1|N)$-supermultiplet with no central charge has an Adinkra is deferred to Ref.\cite{r6-5}, which will also address the (rather more general, but hitherto unemployed) cases when this is not true.

\paragraph{4}
 The counter-example examined above is not isolated, merely the simplest. There exist similar counter-examples for higher $N$, the easiest of which is given by the pair of $(6,8,2)$-dimensional $(1|6)$-supermultiplets corresponding to the Adinkras
\begin{equation}
 \vC{ \begin{picture}(120,45)(0,0)
  \put(-5,3){\includegraphics[height=40mm]{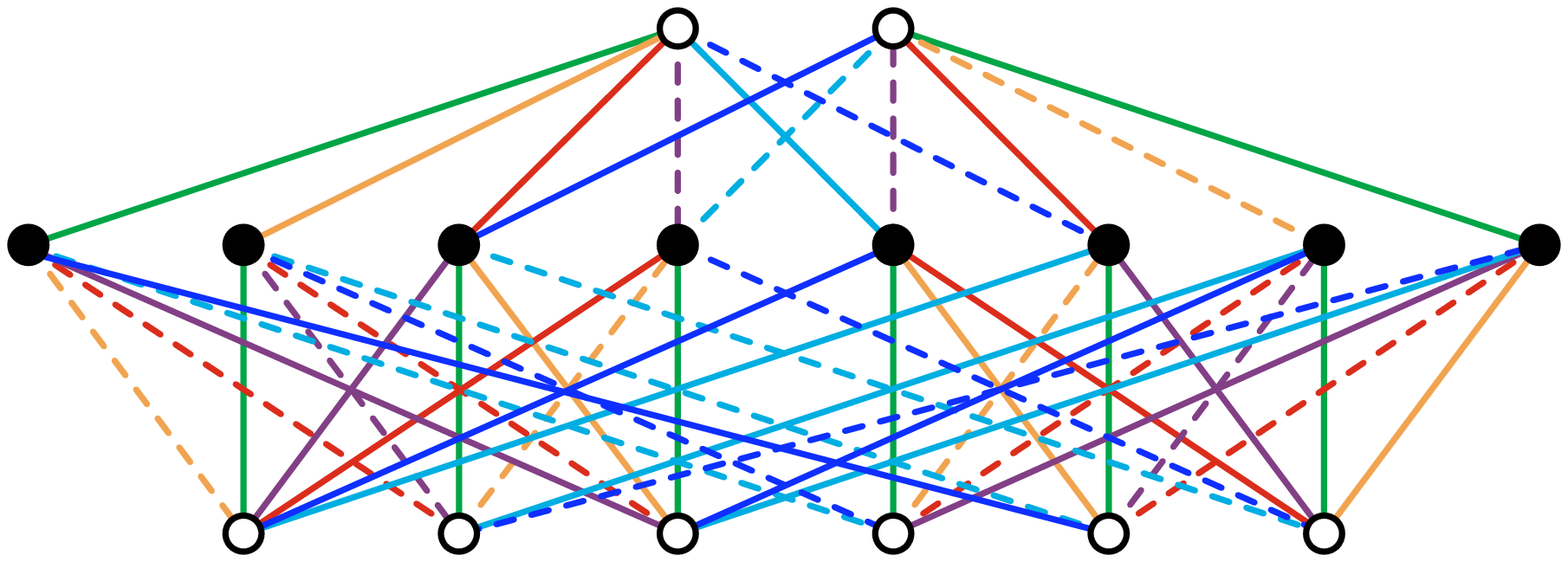}}
 \end{picture}}
 \label{e682x4}
\end{equation}
and
\begin{equation}
 \vC{ \begin{picture}(120,45)(0,0)
  \put(-5,3){\includegraphics[height=40mm]{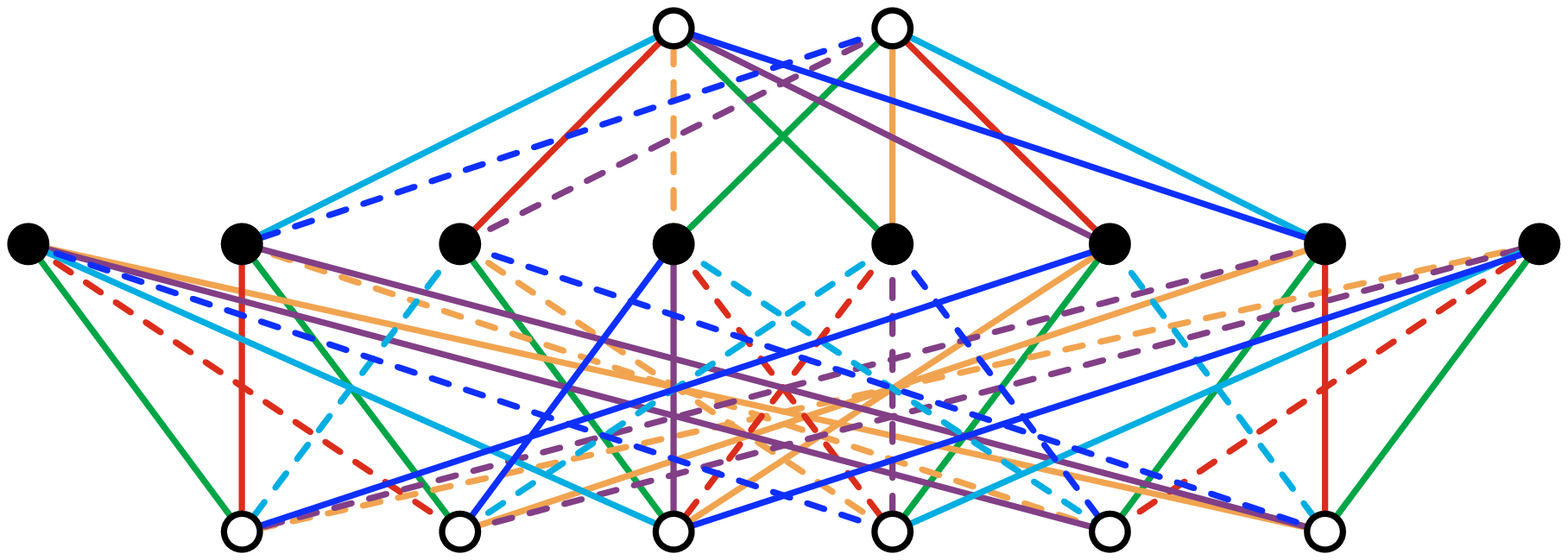}}
 \end{picture}}
 \label{e682x6}
\end{equation}
Clearly, these extend the pair\eqs{e682w2}{e682w4} by adding a $6^{th}$ supersymmetry (dark blue edges) while maintaining the distinction between the two. On the other hand, by deletion of either the green or the orange edges turns\eq{e682x4} into\eq{e682w4}, while the deletion of the edges of any other one color turns\eq{e682x4} into\eq{e682w2}, in both cases up to some re-coloring and parity, \ie, dashedness changes of the edges.

\appendix
\section{Supersymmetry Transformation Rules}
 \label{ABC}
For the benefit of the Reader familiar with the standard component field notation, we read off the supersymmetry transformation rules from the Adinkra\eq{e682w2}:
\begin{align}
 \d_Q(\e)\,\f_1
  &= -\TE\j_1 +\GE\j_2 +\PE\j_3 +\RE\j_4 +\CE\j_6~,\label{f1}\\
 \d_Q(\e)\,\f_2
  &= -\TE\j_4 +\GE\j_3 -\PE\j_2 -\RE\j_1 +\CE\j_7~,\label{f2}\\
 \d_Q(\e)\,\f_3
  &= +\TE\j_3 +\GE\j_4 +\PE\j_1 -\RE\j_2 +\CE\j_8~,\label{f3}\\
 \d_Q(\e)\,\f_4
  &= -\TE\j_6 +\GE\j_5 +\PE\j_8 -\RE\j_7 -\CE\j_1~,\label{f4}\\
 \d_Q(\e)\,\f_5
  &= +\TE\j_5 +\GE\j_6 -\PE\j_7 -\RE\j_8 -\CE\j_2~,\label{f5}\\
 \d_Q(\e)\,\f_6
  &= +\TE\j_8 +\GE\j_7 +\PE\j_6 +\RE\j_5 -\CE\j_3~,\label{f6}\\[2mm]
 \d_Q(\e)\,\j_1
  &= -i\TE\df_1 +i\GE F_7 +i\PE\df_3 -i\RE\df_2 -i\CE\df_4~,\label{j1}\\
 \d_Q(\e)\,\j_2
  &= +i\TE F_7 +i\GE\df_1 -i\PE\df_2 -i\RE\df_3 -i\CE\df_5~,\label{j2}\\
 \d_Q(\e)\,\j_3
  &= +i\TE\df_3 +i\GE\df_2 +i\PE\df_1 +i\RE F_7 -i\CE\df_6~,\label{j3}\\
 \d_Q(\e)\,\j_4
  &= -i\TE\df_2 +i\GE\df_3 -i\PE F_7 +i\RE\df_1 -i\CE F_8~,\label{j4}\\
 \d_Q(\e)\,\j_5
  &= +i\TE\df_5 +i\GE\df_4 -i\PE F_8 +i\RE\df_6 +i\CE F_7~,\label{j5}\\
 \d_Q(\e)\,\j_6
  &= -i\TE\df_4 +i\GE\df_5 +i\PE\df_6 +i\RE F_8 +i\CE\df_1~,\label{j6}\\
 \d_Q(\e)\,\j_7
  &= -i\TE F_8 +i\GE\df_6 -i\PE\df_5 -i\RE\df_4 +i\CE\df_2~,\label{j7}\\
 \d_Q(\e)\,\j_8
  &= +i\TE\df_6 +i\GE F_8 +i\PE\df_4 -i\RE\df_5 +i\CE\df_3~,\label{j8}\\[2mm]
 \d_Q(\e)\,F_7
  &= +\TE\dj_2 +\GE\dj_1 -\PE\dj_4 +\RE\dj_3 +\CE\dj_5~,\label{f7}\\
 \d_Q(\e)\, F_8
  &= -\TE\dj_7 +\GE\dj_8 -\PE\dj_5 +\RE\dj_6 -\CE\dj_4~,\label{f8}
\end{align}
where we have labeled the supersymmetry parameters, $\e^I$, redundantly: by their numeric superscript and also by the corresponding color; the latter matches the colors of the corresponding edges\eqs{e682w2}{e682w4} to facilitate comparison. The engineering dimensions of the component fields:
\begin{equation}
 [\f_I]\strut_{I=1,\cdots,6}~=~\big([\j_I]\strut_{I=1,\cdots,8}-\inv2\big)
  ~=~\big([F_7]-1\big)~=~\big([ F_8]-1\big)~,
\end{equation}
and the nodes in the Adinkras\eqs{e682w2}{e682w4} are stacked at heights that reflect this.

Similarly, we read off the supersymmetry transformation rules from the Adinkra\eq{e682w4}:
\begin{align}
 \d_Q(\e)\,\f_1
  &= -\TE\j_8 +\GE\j_1 -\PE\j_7 +\RE\j_2 -\CE\j_3~,\label{F1}\\
 \d_Q(\e)\,\f_2
  &= \TE\j_7 +\GE\j_2 -\PE\j_8 -\RE\j_1 -\CE\j_5~,\label{F2}\\
 \d_Q(\e)\,\f_3
  &= +\TE\j_6 +\GE\j_3 +\PE\j_4 -\RE\j_5 +\CE\j_1~,\label{F3}\\
 \d_Q(\e)\,\f_4
  &= -\TE\j_3 +\GE\j_6 -\PE\j_5 -\RE\j_4 +\CE\j_8~,\label{F4}\\
 \d_Q(\e)\,\f_5
  &= -\TE\j_2 +\GE\j_7 +\PE\j_1 -\RE\j_8 -\CE\j_4~,\label{F5}\\
 \d_Q(\e)\,\f_6
  &= +\TE\j_1 +\GE\j_8 +\PE\j_2 +\RE\j_7 -\CE\j_6~,\label{F6}\\[2mm]
 \d_Q(\e)\,\j_1
  &= i\TE\df_6 +i\GE\df_1 +i\PE\df_5 -i\RE\df_2 +i\CE\df_3~,\label{J1}\\
 \d_Q(\e)\,\j_2
  &= -i\TE\df_5 +i\GE\df_2 +i\PE\df_6 +i\RE\df_1 +i\CE F_7~,\label{J2}\\
 \d_Q(\e)\,\j_3
  &= -i\TE\df_4 +i\GE\df_3 -i\PE F_8 +i\RE F_7 -i\CE\df_1~,\label{J3}\\
 \d_Q(\e)\,\j_4
  &= -i\TE F_7 +i\GE F_8 +i\PE\df_3 -i\RE\df_4 -i\CE\df_5~,\label{J4}\\
 \d_Q(\e)\,\j_5
  &= +i\TE F_8 +i\GE F_7 -i\PE\df_4 -i\RE\df_3 -i\CE\df_2~,\label{J5}\\
 \d_Q(\e)\,\j_6
  &= +i\TE\df_3 +i\GE\df_4 +i\PE F_7 +i\RE F_8 -i\CE\df_6~,\label{J6}\\
 \d_Q(\e)\,\j_7
  &= +i\TE\df_2 +i\GE\df_5 -i\PE\df_1 -i\RE\df_6 +i\CE F_8~,\label{J7}\\
 \d_Q(\e)\,\j_8
  &= -i\TE\df_1 +i\GE\df_6 -i\PE\df_2 -i\RE\df_5 +i\CE\df_4~,\label{J8}\\[2mm]
 \d_Q(\e)\,F_7
  &= -\TE\dj_4 +\GE\dj_5 +\PE\dj_6 +\RE\dj_3 +\CE\dj_2~,\label{F7}\\
 \d_Q(\e)\, F_8
  &= +\TE\dj_5 +\GE\dj_4 -\PE\dj_3 +\RE\dj_6 +\CE\dj_7~,\label{F8}
\end{align}
The differences between the Adinkras\eq{e682w2} and\eq{e682w4} are easier to spot than those between the system\eqs{f1}{f8} and the system\eqs{F1}{F8}. Notice, however, that the transformation of $\j_1,\j_8$ in\eq{J1} and\eq{J8} are total derivatives, unlike any of \eqs{j1}{j8}.

\bigskip\paragraph{\sf\bfseries Note added}\hspace{-3mm}(April 16, 2007): After posting of this article on the {\tt arXiv.org} (Nov.~6, 2006), the Authors of Refs.\cite{rKRT} disagreed with our assessment of their publication; to clarify, we have now added footnote~\ref{f:Exp}.
 Independently, two of the Authors of Refs.\cite{rKRT} produced Ref.\cite{rKT07}. Herein, they ``refine'' what Refs.\cite{rKRT} claimed to have been a ``complete classification'', by listing supermultiplets of different ``connectivity''---a term used herein first, to convey the more precise notion of {\em topology\/}\cite{r6-1}. This ``refinement'' turns out to corroborate the inequivalence of\eq{e682w2} and\eq{e682w4} and so our claim that the classification of Ref.\cite{rKRT} was incomplete.
 
 Ref.\cite{rKT07} in turn purports to have found counter-examples to Theorem~4.1 and Corollary~4.2 from Ref.\cite{r6-1} and presents explicitly one such distinct pair, in Eqs.~(4.13) and~(4.14), ``while admitting the same {\em number\/} of sources and the same {\em number\/} of targets''~{\small[{\em italics\/} ours]}. As explained in the penultimate paragraph of section~\ref{s:CE} herein, Theorem~4.1 and Corollary~4.2 from Ref.\cite{r6-1} claims not the {\em number\/} but the {\em sets\/} of sources or sinks\Ft{Being a {\em subgraph\/} of the Adinkra, they are specified by their connection to the rest of the Adinkra.} to specify a supermultiplet uniquely. Examples~(4.13) and~(4.14) of Ref.\cite{rKT07} are therefore counter-examples to those Author's misquoting of Ref.\cite{r6-1}, not Ref.\cite{r6-1} itself. The Adinkras corresponding to the purported counter-examples are:
\begin{equation}
 \vC{\includegraphics[height=40mm]{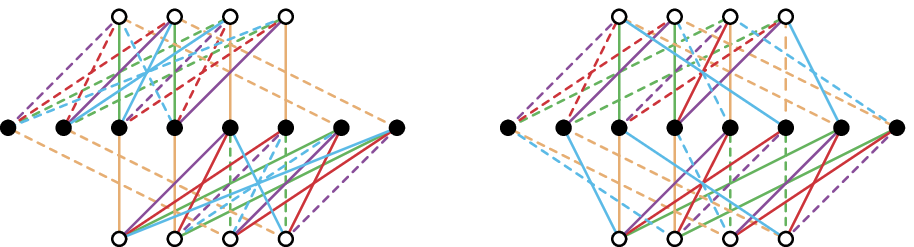}}
 \label{eKTs}
\end{equation}
and are obviously different; the reader is invited to compare the ease of this distinction to the comparison of the explicit $16+16$ transformation equations in Ref.\cite{rKT07}. Furthermore, the two Adinkras\eq{eKTs} are indeed distinguished by their {\em sets\/} of sources/targets---not solely the {\em number\/} of them, as Ref.\cite{rKT07} misquotes Ref.\cite{r6-1}. In both cases, the sets of sources consist of the four bosons in the bottom layer: the distinction is clearly displayed not by their number, but by the different connectivity to the rest: In the left-hand side Adinkra, the four source-bosons connect to four of the fermions by a single edge, and by four edges to the other four fermions; not so in the right-hand side one. No field redefinition can erase this {\em topological\/} distinction.
 
 Finally, while Ref.\cite{rKT07} produces ``the {\em unique\/} pair of $N=5$ irreps (the $(4,8,4)_A$ and the $(4,8,4)_B$ multiplets) differing by connectivity, while admitting the same number of sources and the same number of targets,''~{\small[{\em italics\/} ours]} there in fact are more\Ft{In fact, there can exist additional distinct supermultiplets, obtained by changing the sign of a select subset of the $Q_I$'s, represented here by dashed {\it vs}.\ solid edges.}:
\begin{equation}
 \vC{ \begin{picture}(140,90)(0,0)
  \put(0,0){\includegraphics[height=90mm]{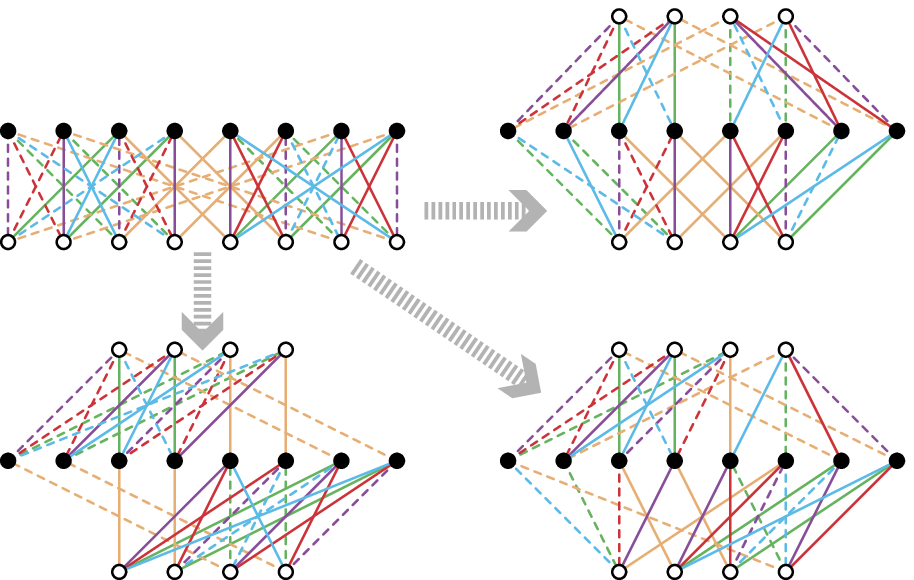}}
   \put(18,45){$4_L{+}0_R$}
   \put(59,35){$3_L{+}1_R$}
   \put(67,61){$2_L{+}2_R$}
 \end{picture}}
 \label{eNonKT}
\end{equation}
All three of the resulting Adinkras have the same {\em number\/} of sources (four), but the connections of these to the rest distinguish one from another. The lower-left and the top-right Adinkra are up to some recoloring and horizontal repositioning the same as\eq{eKTs}, and correspond to the examples~(4.13) and~(4.14) of Ref.\cite{rKT07}; the supermultiplet of the lower-right Adinkra is missed by Ref.\cite{rKT07}.

 The ease with which the missing $(4,8,4)$-dimensional $(1|5)$-supermultiplet found using Adinkras stems from examining the inequivalent ways in which we can raise four white nodes of the $(8,8)$-dimensional representation with the Adinkra in the top-left corner of\eq{eNonKT}.
 The nodes comprising the left half of this Adinkra are clearly distinguished from those in the right half.
 To obtain a $(4,8,4)$-dimensional representation, we can raise:
 \begin{enumerate}\itemsep=-3pt\vspace{-2mm}
 \item $4_L{+}0_R$: 4 white nodes from the left half, producing the left-hand side Adinkra in\eq{eKTs}, corresponding to example~(4.13) of Ref.\cite{rKT07};
 \item $2_L{+}2_R$: 2 white nodes from the left and two from the right half, producing the right-hand side Adinkra in\eq{eKTs}, corresponding to example~(4.14) of Ref.\cite{rKT07};
 \item $3_R{+}1_L$: 3 from the left, 1 from the right half, producing\eq{eNonKT}, which Ref.\cite{rKT07} missed.
\end{enumerate}\vspace{-1mm}
We should trust the Reader to have been convinced by now of the ease of use of Adinkras in listing inequivalent representations of $(1|N)$-supersymmetry, akin to Feynman diagrams in field theory.

\bigskip\bigskip\paragraph{\sf\bfseries Acknowledgments:}
 The research of S.J.G.\ is supported in part by the National Science Foundation Grant PHY-0354401, the endowment of the John S.~Toll Professorship and the CSPT.
 T.H.\ is indebted to the generous support of the Department of Energy through the grant DE-FG02-94ER-40854.
 The Adinkras were drawn with the aid of {\em Adinkramat\/}~\copyright\,2006 by G.~Landweber.

\bibliographystyle{habbrv}
\bibliography{Refs}
\end{document}